\documentclass[11pt]{llncs}
\pdfoutput=1 
\usepackage{wrapfig}
\usepackage{comment}
\usepackage{ifthen}
\usepackage{subfig}
\usepackage{graphicx}
\usepackage{latexsym}
\usepackage{amssymb,amsmath}
\usepackage{color}
\usepackage{comment}
\usepackage{tabularx}
\usepackage[lined, ruled, linesnumbered, noresetcount,noend,oldcommands]{algorithm2e}
\usepackage{multicol}
\dontprintsemicolon\usepackage{authblk}
\usepackage{enumitem}

\newcommand{\thesystem}{\textsc{Lilac-TM}\xspace}

\newcommand{\lref}[1]{line~\nlSty{\ref{#1}}}

\newcommand{\llref}[2]{lines~\nlSty{\ref{#1}}--\nlSty{\ref{#2}}}

\newboolean{showcomments}
\setboolean{showcomments}{false}
\ifthenelse{\boolean{showcomments}}
{ \newcommand{\mynote}[3]{
    \fbox{\bfseries\sffamily\scriptsize#1}
    {\small$\blacktriangleright$\textsf{\emph{\color{#3}{#2}}}$\blacktriangleleft$}}}
{ \newcommand{\mynote}[3]{}}

\begin{document}

\title{Exploiting Locality in Lease-Based Replicated\\ Transactional Memory via Task Migration}

\author{Danny Hendler\inst{1}, Alex Naiman\inst{1}, Sebastiano Peluso\inst{2}, Francesco Quaglia\inst{2}, Paolo Romano\inst{3} \and Adi Suissa\inst{1}}
\institute{Ben-Gurion University of the Negev, Israel
\and Sapienza University of Rome, Italy
\and Instituto Superior T{\'e}cnico, Universidade de Lisboa/INESC-ID, Portugal}



\maketitle

\begin{abstract}
We present \emph{Lilac-TM}, the first locality-aware Distributed Software Transactional Memory (DSTM) implementation. Lilac-TM is a fully decentralized lease-based replicated DSTM. It employs a novel self-optimizing lease circulation scheme based on the idea of dynamically determining whether to migrate transactions to the nodes that own the leases required for their validation, or to demand the acquisition of these leases by the node that originated the transaction.
Our experimental evaluation establishes that Lilac-TM provides significant performance gains for distributed workloads exhibiting data locality, while typically incurring little or no overhead for non-data local workloads.

\end{abstract}

	\section{Introduction}
\label{sec:intro}

\sloppy Transactional Memory (TM) has emerged as a promising programming paradigm for concurrent applications, which
provides a programmer-friendly alternative to traditional lock-based synchronization.
Intense research work on both software and hardware TM approaches \cite{tl2,guerraouiOpacity,stm-lan-support,HTM-93,Shavit1995} and the inclusion  of TM support in world-leading multiprocessor hardware and open source compilers \cite{inteloptimize,GCC} extended the traction it had gained in the research community also to the mainstream software industry.

\emph{Distributed Software TM} (DSTM) systems extend the reach of the TM model to distributed applications.
An important lesson learnt by the deployment of the first enterprise-class TM-based applications \cite{CachopoThesis,ladis08} is that in order to permit scalability 
and meet the reliability requirements of real-world applications, DSTMs must support data replication. As a result, several replication techniques for distributed TM have been proposed, deployed over a set of shared-nothing multi-core systems~\cite{Sinfonia,fortress,prdc09,SaadR12,DiSTM}, as typical of cloud computing environments.

A key challenge faced by replicated DSTMs, when compared, for instance, with more conventional transactional systems (such as relational databases), is the large increase of the communication-to-computation ratio~\cite{ladis08}: unlike classical DBMSs, DSTMs avoid disk-based logging and rely on in-memory data replication to achieve durability and fault-tolerance; further, the nature of the programming interfaces exposed by DSTMs 
drastically reduces the latencies of accessing/manipulating data, leading to significant reduction of the duration of typical TM transactions (again, when compared to typical database transactions). Overall, the reduction of the transaction processing time results in the growth of the relative cost of the distributed (consensus-based \cite{luisBook}) coordination activities required by conventional replication protocols, and in a corresponding increase of their relative overhead.

\subsubsection*{Model and Background}
\label{background}

We consider a classical asynchronous distributed system
model\,\cite{luisBook} consisting of a set of processes $\Pi =
\{p_1,\ldots,p_n\}$ that communicate via message passing and can fail
according to the fail-stop (crash) model. We assume that a majority of
processes is correct and that the system ensures sufficient
synchrony for implementing a \emph{View Synchronous Group
Communication Service} (GCS)\,\cite{gcs-keidar}.

GCS provides two complementary services: group membership and multicast communication.
Informally, the role of the \emph{group membership service} is to provide each
participant in a distributed computation with information about which process is
active (or reachable) and which is failed (or unreachable). Such
information is called a \emph{view} of the group of participants.
We assume that the GCS provides a \emph{view-synchronous primary-component group membership service} \cite{baldoni-primary}, which maintains a single agreed view of the group at any given time and provides processes with information on whether they belong to the  primary component.

The \emph{multicast communication service} allows a member to send messages with different reliability and ordering properties to the group of participants. We assume the availability of two communication services: \emph{Optimistic Atomic Broadcast} (OAB)\,\cite{atomicBroadcast} and \emph{Uniform Reliable
Broadcast} (URB)\,\cite{luisBook}. URB is defined by the primitives {\em UR-broadcast}($m$) and {\em UR-deliver}($m$) that guaranty causal order message delivery. Three primitives define
OAB: {\em OA-broadcast}($m$), which is used to broadcast message $m$;
{\em Opt-deliver}($m$), which delivers message $m$ without providing
ordering guarantees; {\em TO-deliver}($m$), which delivers message $m$
in the final total order.




The ALC (Asynchronous Lease Certification) protocol (fully described in \cite{CarvalhoRR10}) is based on the \emph{lease} concept. A \emph{lease} is an ownership token that grants a node temporary privileges on the management of a subset of the replicated data-set. ALC associates leases with data items indirectly through \emph{conflict classes}, each of which may represent a set of data items. This allows flexible control of the granularity of the leases abstraction, trading off accuracy (i.e.,~avoidance of aliasing problems) for efficiency (amount of information exchanged among nodes and maintained in-memory) \cite{tradeoff}.

With ALC, a transaction is executed based on local data, avoiding any inter-replica synchronization until it enters its commit phase. At this stage, ALC acquires a lease for the transaction's accessed data items, before proceeding to validate the transaction. In case a transaction $T$ is found to have accessed stale data, $T$ is re-executed without releasing acquired leases. This ensures that, during $T$'s re-execution, no other replica can update any of the data items accessed during $T$'s first execution, which guarantees the absence of remote conflicts on the subsequent re-execution of $T$, provided that the same set of conflict classes accessed during $T$'s first execution is accessed again.


To establish lease ownership, ALC employs the OAB communication service. Disseminating data items of committed transactions and lease-release messages in done using the URB service.
The ownership of a lease ensures that no other replica will be allowed to successfully validate any conflicting transaction, making it unnecessary to enforce distributed agreement on the global serialization order of transactions.
ALC takes advantage of this by limiting the use of atomic broadcast exclusively for establishing the lease ownership. Subsequently, as long as the lease is owned by the replica, transactions can be locally validated and their updates can be disseminated using URB, which can be implemented in a much more efficient manner than OAB.

\subsubsection*{Our Contributions}
In this paper,
we present an innovative, fully decentralized, LocalIty-aware LeAse-based repliCated TM (\thesystem). 
\thesystem aims to maximize system throughput
via a distributed self-optimizing lease circulation scheme based on the idea of dynamically determining whether to migrate transactions to the nodes that own the leases required for their validation, or to demand the acquisition of these leases by the transaction originating node.

\thesystem's flexibility in deciding whether to migrate data or transactions allows it not only to take advantage of the data locality present in many application workloads, but also to further enhance it by turning a node $N$ that frequently accesses a set of data items $D$ into an attractor for transactions that access subsets of $D$ (and that could be committed by $N$ avoiding any lease circulation). This allows \thesystem to provide two key benefits: (1) limiting the frequency of lease circulation,
and (2) enhancing contention management efficiency. 
In fact, with \thesystem, conflicting concurrent transactions have a significantly higher probability of being  executed on the same node, which prevents them from
 incurring the high costs of distributed conflicts. This paper makes the following contributions:

\begin{enumerate}
\item 
We present a fully-fledged prototype of \thesystem, which implements a replicated DSTM based on Java technology.
\item  ALC generates \emph{a single} lease request for the entire transaction data-set.
This limits the exploitation of data-locality, since another local transaction
may reuse the lease only if its data-set is a subset of another lease owned by the node.
To allow efficient exploitation of data-locality by \thesystem, we present a new version of ALC that supports fine-grained leases (FGL-ALC), which,
 instead of acquiring one lease for the entire data-set, acquires a set of leases, one per item of the data-set.
\item  We conduct a comprehensive comparative performance analysis, evaluating the performance gains obtained by the new locality-aware infrastructure and algorithms we developed in comparison with ALC.
    Our results establish that replacing ALC by FGL-ALC yields significant performance boost for workloads possessing data locality. When Lilac-TM is used on top of the new lease-management infrastructure, performance gains are greatly increased, providing up to 3.2 times the throughput of the baseline implementation.
\end{enumerate}






\section{Related Work}
\label{related-work}


Literature results on replication of transactional systems provide solutions based on protocol
specifications (see, e.g., \cite{dangers,KemmeA00,DBSM}), replication-middleware designs (see, e.g., \cite{LinKPJ05,Patino-MartinezJKA05,PedoneF08}) and extensions of the inner logic of transactional systems in order to support specific replication strategies (see, e.g., \cite{KemmeA00,WuK05}).
As shown in \cite{WiesmannS05}, the approaches exploiting (Optimistic) Atomic Broadcast (OAB)
\cite{pedone-optimistic} for replica coordination seem the most promising, which motivated
 recent proposals that specifically target replication of (S)TM systems
to heavily rely on (O)AB-based
distributed coordination \cite{nca2010,PalmieriQR11}. In particular, these approaches exploit combined usage of OAB
and speculative execution of optimistically (early) delivered transactional requests
in order to effectively overlap computation and coordination phases.
On the other hand, the above approaches have been designed and studied for the case of \emph{active replication}, where update-transactions are broadcast and  processed at all the replicated sites.
Compared with these approaches, our proposal still aims at (partially) removing the cost of coordination from the critical-path of transaction processing (since transactions can be executed and safely committed
with no preventive coordination action in case the node executing them already holds the requested leases). However, any update transaction is run locally at an individual site (its updates are then propagated to the remote sites), which permits better resource exploitation for update-intensive workloads.

As for the avoidance of running update-transactions at all the sites, several approaches exist (either AB-based~\cite{DBSM,alonso98,prdc09}, or 2PC-based~\cite{PelusoRRQR12,SchiperSP10,PelusoRQ12})
which exploit optimistic processing schemes coupled with globally ordered commit-time certification.
However, these approaches may suffer from non-optimal transaction-abort rates in case of high-conflict workloads, since a (re-run) transaction may be aborted multiple times due to repeated conflicts with remotely running transactions. This is avoided by our proposal since, once the leases on the data requested by the transaction have been acquired, the transaction can be committed if no local conflicts are detected. Further, as already discussed, by re-using a lease across multiple transactions, the lease acquisition costs (i.e.,~a distributed consensus) can be significantly amortized.

The approach in \cite{Sinfonia} allows reducing the impact of the 2PC coordination protocol by assuming that transactions' data-set and transactional operations are known in advance.
In contrast, our proposal does not require a-priori knowledge of the set of data to be read/written by transactions, and hence is suited for more general programming paradigms.



Exploitation of the access locality in distributed/replicated (S)TM systems has also been pursued by proposals relying on data-flow \cite{SaadR12,KimR10}. This is done via optimized scheduling policies for the transactions (depending on the target data set) and via
optimized move of the ownership of data-slices within the data-flow model (still in relation to the locality of the accesses). Our approach has relations with these proposals in that leases are analogous to the ownership concept (and allow the lease-owner node to autonomously and safely take commit decisions for transactions only touching the leased data-slices). However, the above works do not entail any form of transaction migration as in our approach.

\section{\thesystem}
\label{lilac-tm}


\begin{figure}
	\centering
	\includegraphics[scale=0.42, angle=0]{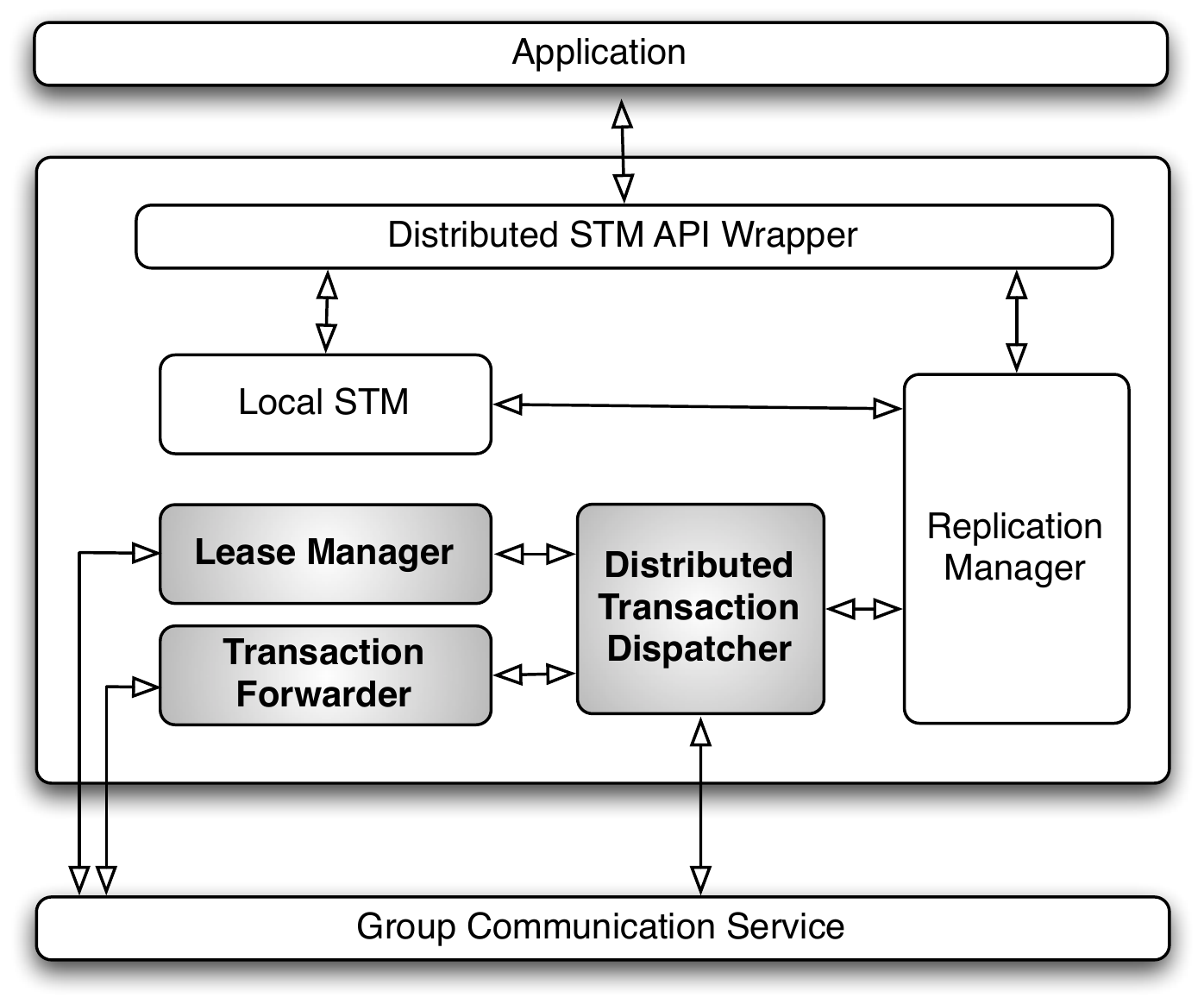}
	\captionsetup{textfont=small}
	\caption{Middleware architecture of a \thesystem replica.}
	\label{fig:architecture}
\end{figure}



Figure \ref{fig:architecture} provides an overview of the software architecture of each replica of \thesystem, highlighting in gray the modules that were either re-designed or that were not originally present in ALC.

The top layer is a wrapper that
intercepts application level calls for transaction demarcation
(i.e., to begin, commit or abort transactions), without interfering
with application accesses (read/write) to the transactional data
items, which are managed directly by the underlying local STM layer. This
approach allows transparent extension of the classic STM programming model to a distributed setting.

The prototype of \thesystem has been built by extending the ALC implementation shipped in the GenRSTM framework~\cite{genrstm}. GenRSTM has been designed to support, in a modular fashion, a range of heterogeneous algorithms across the various layers of the software stack of a replicated STM platform.
\thesystem inherits this flexibility from GenRSTM. In this work, we use TL2~\cite{tl2} as the local STM layer.

The Replication Manager (RM) is the component in charge of interfacing the local STM layer with its replicas deployed on other system nodes. The RM is responsible of coordinating the commit phase of both remote and local transactions by:
(i) intercepting commit-request events generated by local transactions and triggering a distributed coordination phase aimed at determining transactions' global serialization order and detecting the presence of conflicts with concurrently executing remote transactions; and
(ii) validating remote transactions and, upon successful validation, committing them by atomically applying their write-sets in the local STM.

At the bottom layer we find a GCS (Appia~\cite{appia} in our prototype), which, as mentioned in Section \ref{sec:intro}, provides the view synchronous membership, OAB and URB services.

The role of the Lease Manager (LM) is to ensure that no two replicas simultaneously disseminate updates for conflicting transactions. To this end, the LM exposes an interface consisting of two methods, \textsc{getLease}() and \textsc{finishedXact}(), which are used by the RM to acquire/release leases on a set of data items. This component was originally introduced in ALC and has been re-designed in this work to support \emph{fine-grained leases}. As we explain in more detail in Section \ref{sec:lor}, fine-grained leases facilitate the exploitation of locality and consequently may provide a significant reduction of lease circulation.

\thesystem includes two new modules that were not present in the original ALC system. These are the Transaction Forwarder (TF) and the Distributed Transaction Dispatcher (DTD).

As the name suggest, the TF is responsible for managing the forwarding of a transaction to a different node in the system. The transaction forwarding mechanism employed by \thesystem represents an alternative mechanism to the lease-based certification scheme introduced in ALC. Essentially, both transaction forwarding and lease-based replication strive to achieve the same goal: minimizing the execution rate of expensive Atomic Broadcast-based consensus protocols to determine the outcome of commit requests.
ALC's lease mechanism pursues this objective by allowing a node that owns the leases associated with its dataset to validate transactions and disseminate their writesets without executing consensus protocols. Still, as described in Section \ref{sec:intro}, acquiring a lease remains an expensive operation, as it requires the execution of a consensus protocol, encapsulated by the OAB service.

The transaction forwarding scheme introduced in this work aims at reducing the frequency of lease requests triggered in the system, by migrating the execution of transactions to remote nodes that may process them more efficiently. This is the case, for instance, if some node $n$ owns the set of leases required to certify and commit a transaction $T$ originated on some remote node $n'$.
In this scenario, in fact, $n$ could validate $T$ locally, and simply disseminate its writeset in case of success. Transaction migration may be beneficial also in subtler scenarios in which, even though no node already owns the leases required to certify a transaction $T$, if $T$'s originating node were to issue a lease request on $T$'s behalf, it would be necessary to revoke leases that are being utilized with high frequency by some other node, say $n''$. In this case, it is preferable to forward the transaction to $n''$ and have $n''$ acquire the lease on behalf of $T$, as this would reduce the frequency of lease circulation and increase throughput in the long term.

The decision of whether to migrate a transaction's execution to another node or to issue a lease request and process it locally is far from being a trivial one. The transaction scheduling policy should take load balancing considerations into account and ensure that the transaction migration logic avoids excessively overloading any subset of nodes in the system.
In \thesystem, the logic for determining how to manage the commit phase of transactions is encapsulated by the DTD module. In this paper we propose two decision policies backed by a precise and efficiently solvable formulation in terms of an Integer Linear Programming optimization problem. The two proposed policies approach the problem from opposite perspectives, by aiming to minimize the short-term versus the long-term costs associated with the handling of transactions' commit phase.

We describe the fine-grained lease management scheme, the TF and the DTD in the following.

\newcommand{\proc}[1]{\textbf{procedure} \textsl{#1} \textbf{do}}
\newcommand{\uponevent}[1]{\textbf{upon} \textsl{#1} \textbf{do}}
\newcommand{\uponbigevent}[2]{\textbf{upon} \textsl{#1}  \\ \textsl{#2} \textbf{do}}
\newcommand{\trigger}[1]{\textbf{trigger} #1}
\newcommand{\setup}[1]{\textbf{setup} \textsl{#1}}
\newcommand{\wait}[1]{\textbf{wait until} \textsl{#1}}
\newcommand{\tab}{\hspace*{2em}}
\newcommand{\remove}[1]{}

\SetKwBlock{finishedTransactionBlock}
	{\texttt{void}
		\textsc{FinishedXact}{(Set$<${\tt LOR}$>$ \textnormal{S})}}
	{end}
	
\SetKwBlock{getLeaseBlock}
	{\texttt{Set$<${\tt LOR}$>$}
		\textsc{GetLease}{({\tt Set} \textnormal{DataSet})}}
	{end}
	
\SetKwBlock{OptdeliverBlock}
	{{\bf upon}
		{\em Opt-deliver}{([{\sf LeaseRequest}, \textnormal{req}]) {\bf from} $p_k$ {\bf do}}}
	{end}	
	
\SetKwBlock{TOdeliverBlock}
	{{\bf upon}
		{\em TO-deliver}{([{\sf LeaseRequest}, \textnormal{req}]) {\bf from} $p_k$ {\bf do}}}
	{end}

\SetKwBlock{URdeliverBlock}
	{{\bf upon}
		{\em UR-deliver}{([{\sf LeaseFreed}, Set$<${\tt LOR}$>$ \textnormal{S}]) {\bf from} $p_k$ {\bf do}}}
	{end}
	
\SetKwBlock{freeLocalLeases}
	{\texttt{void}
		{freeLocalLeases}{(ConflictClass[] CC)}}
	{end}
	
\SetKwBlock{isEnabled}
	{\texttt{boolean}
		{isEnabled}{(Set$<${\tt LOR}$>$ \textnormal{S})}}
	{end}

\subsection{Fine-Grained Leases}
\label{sec:lor}



\sloppy In ALC, a transaction requires a \emph{single lease object}, associated with its data set in its entirety.
A transaction $T$, attempting to commit on a node, may reuse a lease owned by the node only if $T$'s data set is a subset of the lease's items set. Thus, each transaction is tightly coupled with a single lease ownership record. This approach has two disadvantages: i) upon the delivery of a lease request by a remote node that requires even a single data item from a lease owned by the local node, the lease must be released, causing subsequent transactions accessing other items in that lease to issue new lease requests;
ii) if a transaction's data set is a subset of a union of leases owned by the local replica but is not a subset of any of them, a new lease request must be issued. This forces the creation of new lease requests, causing extensive use of {\em TO-broadcast} and increasing commit latency.

To exploit data-locality, {\thesystem} does not use ALC's LM module. Instead, we implemented a new lease manager module that decouples lease requests from the requesting transaction's data set. Rather than having a transaction acquire a single lease encompassing its entire data set, each transaction acquires a set of fine-grained \emph{Lease Ownership Records} (LORs), one per accessed conflict class.

\begin{figure}[t!]
\begin{center}
	\includegraphics[scale=0.25,angle=90]{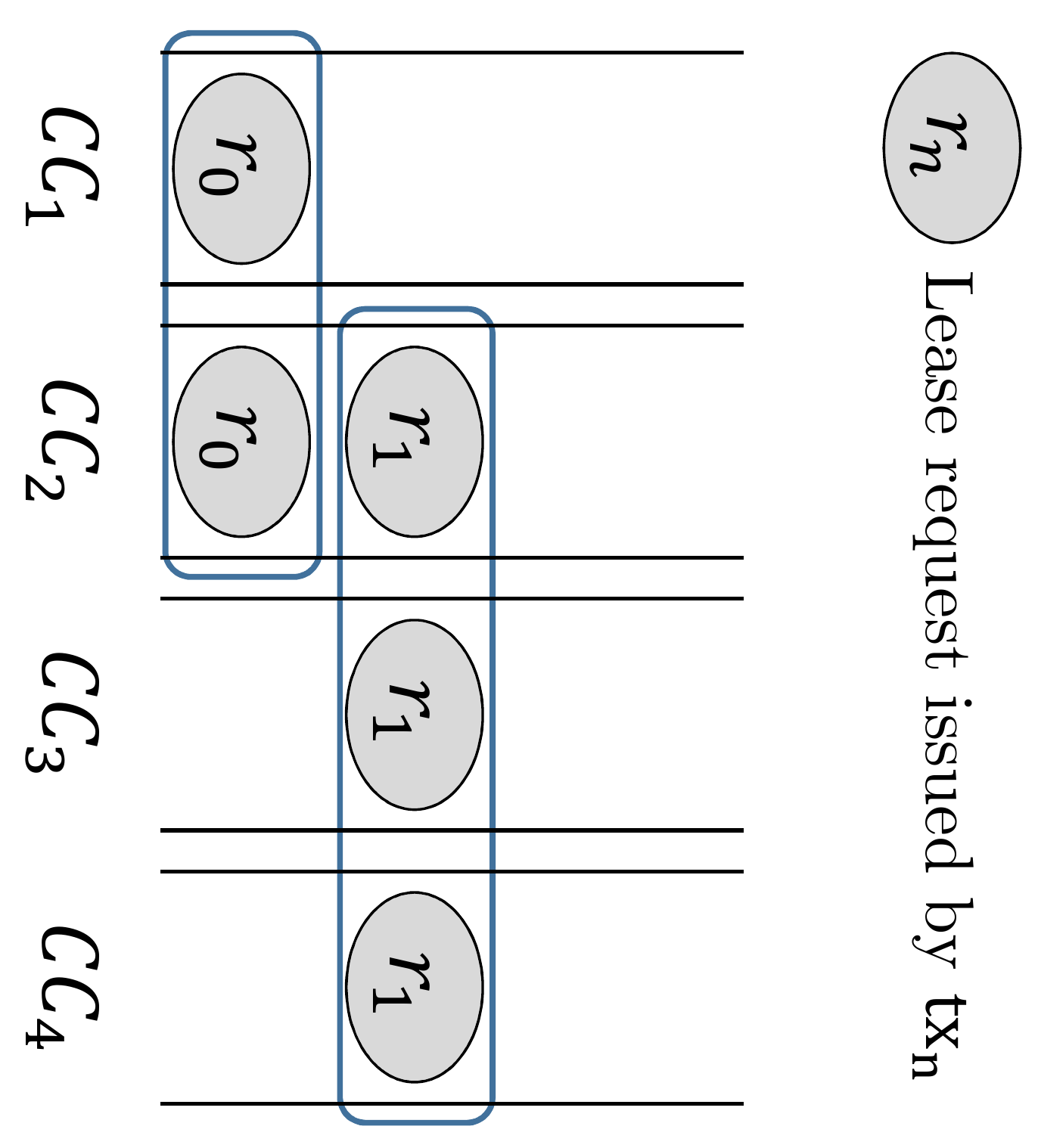}
	\includegraphics[scale=0.25,angle=90]{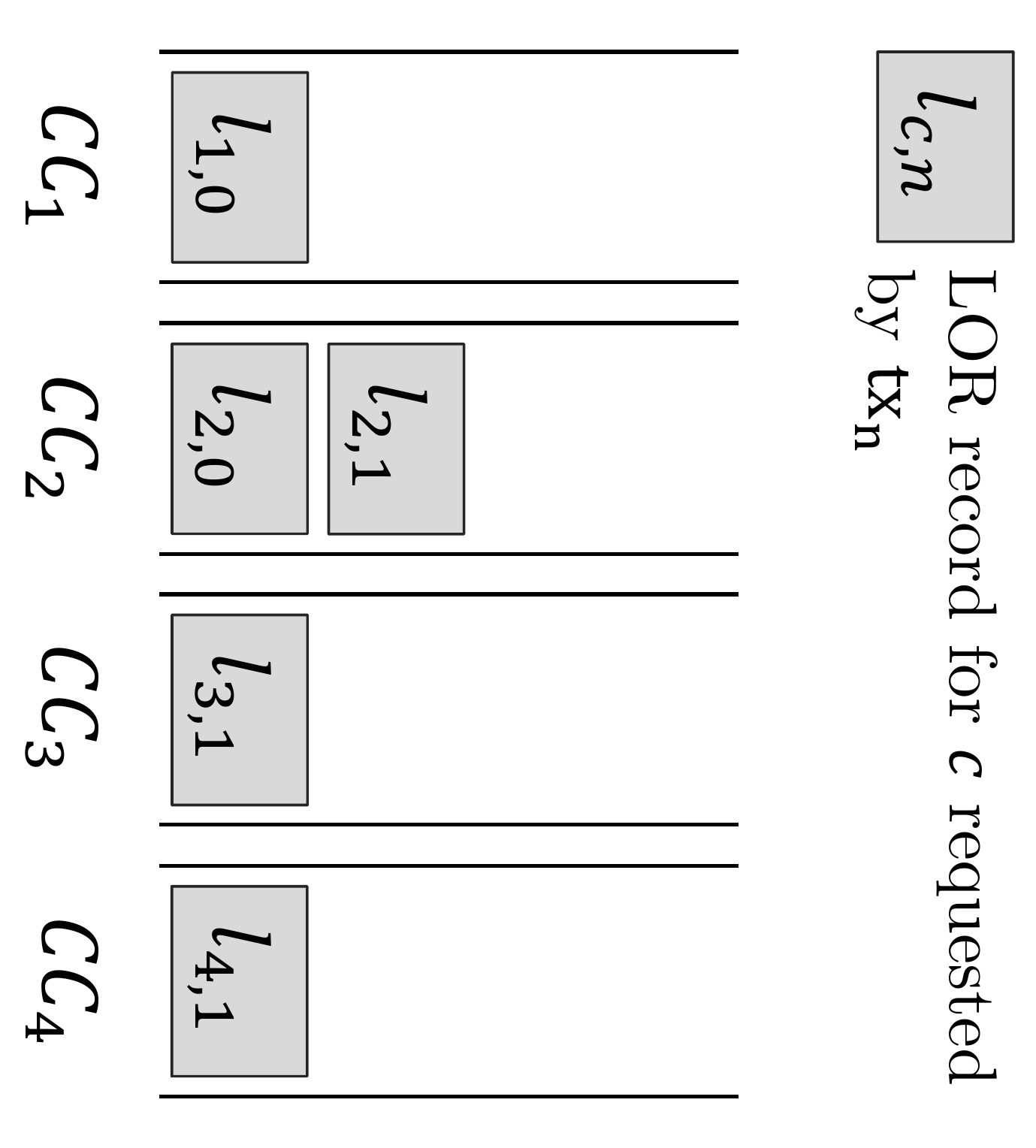}
\end{center}
\captionsetup{textfont=small}
\caption{A view of conflict queues in a replicated memory with 4 conflict classes, after TO-delivering 2 requests. (a) the original ALC protocol, and (b) our new fine-grained leases approach.}
\label{fig:alclors}
\end{figure}

Fig.~\ref{fig:alclors}-(a) and \ref{fig:alclors}-(b) illustrate a scenario in which the new LM benefits from data-locality whereas the ALC LM cannot. They show a replicated system with 4 conflict classes ($CC_1$, $CC_2$, $CC_3$, and $CC_4$) using the original ALC protocol and \thesystem (Fig.~\ref{fig:alclors}(a) and \ref{fig:alclors}(b) respectively).
The figures illustrate the state after TO-delivering two lease requests originating from the same node: the first request is on behalf of transaction $T_0$, accessing $CC_1$ and $CC_2$, followed by a request on behalf of transaction $T_1$, accessing $CC_2$, $CC_3$, $CC_4$. Assume that a third transaction $T_2$, originated on the same node, requires leases on $CC_1$, $CC_3$ and $CC_4$.
With {\thesystem}'s new fine-grained LM, the third request may be granted by ``piggy-backing'' on existing LORs ($l_{1,0}$, $l_{3,1}$, $l_{4,1}$). ALC's LM, on the other hand, must create a new lease request for $CC_1$, $CC_3$ and $CC_4$, generating additional network load and incurring larger latency.

\paragraph{Implementation details.}

ALC's \emph{Replication Manager} (RM) was not changed. It interfaces with the LM via the \textsc{getLease}() and \textsc{finishedXact}() methods for acquiring and releasing leases, respectively. As in ALC, {\thesystem} maintains the indirection level between leases and data items through conflict classes. This allows flexible control of the leases abstraction granularity. We abstract away the mapping between a data item and a conflict class through the \texttt{getConflictClasses}() primitive, taking a set of data items as an input parameter and returning a set of conflict classes.


As in ALC, each replica maintains one main data structure for managing the establishment/release of leases: $CQ$ (Conflict-Queues), an array of FIFO queues, one per conflict class. The CQ keeps track of conflict relations among lease requests of different replicas. Each queue contains LORs, each storing the following data:
(i) \textbf{proc}: the address of the requesting replica;
(ii) \textbf{cc}: the conflict class this LOR is associated with;
(iii) \textbf{activeXacts}: a counter keeping track of the number of active local transactions associated with this LOR, initialized to 1 when the LOR is created; and
(iv) \textbf{blocked} : a flag indicating whether new local transactions can be associated with this LOR - this flag is initialized to false when the LOR is created (in the \texttt{createLorsForConflictClasses} primitive), and set to true as soon as  a remote lease request is received.

\begin{algorithm}[t!]
\begin{multicols}{2}
\scriptsize
\texttt{FIFOQueue}$<${\tt LOR}$>$ CQ[NumConflictClasses]=\{$\perp,\ldots,\perp$\} \\
\BlankLine

\getLeaseBlock{
	ConflictClass[] CC = \texttt{getConflictClasses}(DataSet) \nllabel{getLease:getConfClasses}\\
	\uIf{{\normalfont ($\exists$(Set$<${\tt LOR}$>$)S$\subseteq$CQ s.t. $\forall$cc$\in$CC($\exists$lor$\in$S : (lor.cc=cc $\wedge$ lor.proc=$p_i \wedge \neg$lor.blocked)))} \nllabel{getLease:ifLeasesMayBeReused}}
	{
		\ForEach{\normalfont lor$\in$S \nllabel{getLease:missPiggyStart}}{
			lor.activeXacts$++$\nllabel{getLease:missPiggyEnd}
		}
	} \uElse{\nllabel{getLease:newLeaseStart}
		Set$<${\tt LOR}$>$ S = \texttt{createLorsForConflictClasses}(CC) \\
		LeaseRequest req = {\bf new} \texttt{LeaseRequest}($p_i$,S)\\
{\em OA-broadcast}([{\sf LeaseRequest},req]) \nllabel{getLease:newLeaseEnd}
	}
	\wait{} isEnabled(S) \nllabel{getLease:waitEnabled}\\
	{\bf return} {S} \nllabel{getLease:return}
}
\BlankLine
\finishedTransactionBlock{
	Set$<${\tt LOR}$>$ lorsToFree \\
	\ForEach{\normalfont lor$\in$S}{
		lor.activeXacts$--$ \nllabel{finishedTX:decrement}\\
		\lIf{\normalfont (lor.blocked $\wedge$ lor.activeXacts=0)\nllabel{finishedTX:ifShouldRelease}}{
			lorsToFree=lorsToFree $\cup$ lor
		}
	}
	\lIf{\normalfont (lorsToFree $\neq\emptyset$)}{
	UR-broadcast([{\sf LeaseFreed},lorsToFree]) \nllabel{finishedTX:release}
	}
\nllabel{getLease:FTBEnd}}
\BlankLine
\OptdeliverBlock{
	freeLocalLeases(req.cc)
}
\BlankLine
\TOdeliverBlock{\nllabel{TOdeliver}
	Set$<${\tt LOR}$>$ S = \texttt{createLorsForConflictClasses}(req.cc) \\
	\lForEach{\normalfont lor$\in$S}
		{CQ[lor.cc].\texttt{enqueue}(lor)}
}
\BlankLine
\URdeliverBlock{\nllabel{URdeliver}
	\lForEach{\normalfont lor$\in$S} CQ[lor.cc].\texttt{dequeue}(lor)
}
\BlankLine
\freeLocalLeases{
	Set$<${\tt LOR}$>$ lorsToFree \\
	\ForEach{\normalfont cc $\in$ CC}{
		\If{\normalfont $\exists$lor in CQ[cc] s.t. lor.proc=$p_i$}{
			lor.blocked={\em true} \nllabel{getLease:FLLBlock}\\
			\If{\normalfont (CQ[lor.cc].\texttt{isFirst}(lor) $\wedge$ lor.activeXacts=0)\nllabel{FLL:ifShouldRelease}}{
				lorsToFree=lorsToFree $\cup$ lor
			}
		}
	}
	\lIf{\normalfont (lorsToFree $\neq\emptyset$)}{
UR-broadcast([{\sf LeaseFreed},lorsToFree])\nllabel{FLL:Release}
	}
}
\BlankLine
\isEnabled{
	{\bf return} $\forall$lor$\in$S : CQ[lor.cc].\texttt{isFirst}(lor)
}
\Indm\vspace{1pt}~\\
\end{multicols}
\caption{Lease Manager at process $p_i$}
\label{alg:LOR}
\end{algorithm}
 \renewcommand\baselinestretch{.96}

Algorithm~\ref{alg:LOR} presents the pseudo-code of {\thesystem}'s LM.
The method \textsc{getLease}() is invoked by the RM once a transaction reaches its commit phase. The LM then attempts to acquire leases for all items in the committing transaction's data set.
It first determines, using the \texttt{getConflictClasses}() method, the set of conflict classes associated with the transaction's data set (\lref{getLease:getConfClasses}).
It then checks (in \lref{getLease:ifLeasesMayBeReused}) whether CQ contains a set $S$ of LORs such that i) the LORs were issued by $p_i$, and ii) additional transactions of $p_i$ may still be associated with these LORs (this is the case for each LOR owned by the current node that is not blocked).
If the conditions of \lref{getLease:ifLeasesMayBeReused} are satisfied, the current transaction can be associated with all LORs in $S$ (\llref{getLease:missPiggyStart}{getLease:missPiggyEnd}).
Otherwise, a new lease request, containing the set of LORs, is created and is disseminated using {\em OAB} (\llref{getLease:newLeaseStart}{getLease:newLeaseEnd}). In either case, $p_i$ waits in \lref{getLease:waitEnabled} until $S$ is enabled, that is, until all the LORs in $S$ reach the front of their corresponding FIFO queues (see the \textsc{isEnabled}() method).
Finally, the method returns $S$ and the RM may proceed validating the transaction.

When a transaction terminates,
the RM invokes the \textsc{finishedXact}() method. This method receives a set of LORs and decrements the number of active transactions within each record (\lref{finishedTX:decrement}). Every blocked
LOR that is not used by local transactions is then released by sending a single message via the {\em UR-broadcast} primitive (\llref{finishedTX:ifShouldRelease}{finishedTX:release}).

Upon an {\em Opt-deliver} event of a remote lease request {\em req}, $p_i$ invokes the \textsc{freeLocalLeases}() method, which blocks all LORs owned by $p_i$ that are part of {\em req} by setting their \emph{blocked} field (\lref{getLease:FLLBlock}). Then, all LORs that are blocked and are no longer in use by local transactions are released by sending a single {\em UR-broadcast} message (\llref{FLL:ifShouldRelease}{FLL:Release}). Other LORs required by {\em req} that have local transactions associated with them (if any) will be freed when the local transactions terminate. Blocking LORs is required to ensure the fairness of the lease circulation scheme. In order to prevent a remote process $p_j$ from waiting indefinitely for process $p_i$ to relinquish a lease, $p_i$ is prevented from associating new transactions with existing LORs as soon as a conflicting lease request from $p_j$ is {\em Opt-delivered} at $p_i$ (as described above).

Upon a {\em TO-deliver} event of a lease request {\em req} (\lref{TOdeliver}), $p_i$ creates the corresponding set of LORs, and  enqueues these records in their conflict class queues. The logic associated with a {\em UR-deliver} event (\lref{URdeliver}) removes each LOR specified in the message from its corresponding conflict class queue.

\subsection{Transaction Forwarder}
\label{sec:forwarding}

The Transaction Forwarder is the module in charge of managing the process of migrating transactions between nodes.
As already mentioned, in \thesystem transactions are first locally executed and validated by the node that originated them. Unlike ALC, if the set $S$ of conflict classes accessed by a transaction $T$ is not already owned by its origin node, say $n$, in \thesystem the DTD may decide to avoid requesting leases for $T$, and forward its execution to a different node $n'$ instead. In this case node $n'$ becomes responsible for finalizing the commit phase of the transaction. This includes, first of all, establishing leases on $S$ on behalf of transaction $T$. This can be achieved avoiding any distributed coordination, in case $n'$ already owns all the leases required by $T'$. Otherwise, if some of the leases requested by $T'$ are not owned by $n'$,  $n'$ has to issue lease requests on behalf of $T$ via the OAB service.

Next we can use a remote validation optimization and let $n'$ perform $T$'s final validation upon arrival (without re-executing $T$) in order to detect whether $T$ has conflicts with concurrently committed transactions.\footnote{In order to use this remote validation optimization, the TF module must be augmented with a TM-specific validation procedure and append the appropriate meta-data to forwarding messages. TM-specific adaptation and overhead can be avoided by simply always re-executing the forwarded transaction once it is migrated to $n'$.}
In case of successful validation, $T$ can be simply committed, as in ALC, by disseminating a {\sf Commit} message via the \textit{UR-Broadcast}. Additionally, in \thesystem, this has the effect of unblocking the thread that requested the commit of $T$ on node $n$. On the other hand, if $T$ fails its final validation, it is re-executed on node $n'$ until it can be successfully committed, or it fails for a pre-determined number of attempts.
In this latter case, the origin node is notified of the abort of $T$, and the user application is notified via an explicit exception type.
Note that, in order to commit the transaction associated with the re-execution of $T$, which we denote as $T'$, $n'$ must own the set of conflict classes accessed by $T'$. This may not be necessarily true, as $T'$ and $T$ may access different sets of conflict classes. In this case, \thesystem prevents a transaction from being forwarded an arbitrary number of times, by forcing $n'$ to issue a lease request and acquire ownership of the leases requested by $T'$.

It must be noted that, in order to support the transaction forwarding process, the programming model exposed by \thesystem has to undergo some minor adaptations compared, e.g., with the one typically provided by non-replicated TM systems.
Specifically, \thesystem requires that the transactional code is replicated and encapsulated by an interface that allows to seamlessly re-execute transactions that were originated at different nodes.
 To this end, the transactional logic is wrapped in an object whose attributes encode its input parameters, and which exposes methods supporting its correct serialization and de-serialization, and allowing to trigger the execution of the transactional logic, possibly on a remote node (similarly to RMI). In order to maximize the generality and flexibility of the programming model, the method that supports the execution of a transaction is allowed to return a (typed) result. In case of re-execution on node $n'$ of a transaction $T$ forwarded by node $n$, the transaction's result is piggybacked on the commit message. This allows  to inform the application thread that originated the execution of $T$ on $n$ about  the result generated by $T$ on $n'$ (which may be different from the one originally produced by $T$ on $n$).
 

\subsection{Distributed Transaction Dispatching}
\label{sec:dist}

The DTD module allows encapsulating arbitrary policies to determine whether to process the commit of a transaction locally, by issuing lease requests if required, or to migrate its execution to a remote node. In the following we refer to this problem as the \emph{transaction migration problem}. This problem can be formulated as an Integer Linear Programming (ILP) problem as follows:

\vspace{.2cm}
(1) $\textbf{min}\sum_{i\in \Pi}N_{i} \cdot C(i,S)\label{eq:cost}$ 

(2) $\textnormal{subject to:} \sum_{i\in\Pi}N_{i}=1 \label{eq:eq2}$ 
(3) $CPU_{i}\cdot N_{i} <maxCPU \label{ineq:eq3}$
\vspace{.2cm}


The above problem formulation aims at determining an assignment of the binary vector $N$ (whose entries are all equal to 0 except for one,  whose index specifies the selected node) minimizing a generic cost function $C(i,S)$
that
expresses the cost for node $i$ to be selected for managing the commit phase of a transaction accessing the conflict classes in the set $S$. The notation used in the above constraint problem and in the functions defined in the following is
summarized in Table~\ref{table-LP}.

The optimization problem specifies two constraints. Constraint (2) expresses the fact that a transaction can be certified by exactly a single node in $\Pi$.
Constraint (3) is used to avoid load imbalance between nodes. It states that a node $i$ should be considered eligible for re-scheduling only if its CPU utilization ($CPU_i$) is below a maximum threshold ($maxCPU$).

We now derive two different policies for satisfying the above ILP formulation, which are designed to minimize the long-term and the short-term impact of the decision on how to handle a transaction. We start by defining the cost function $LC(i,S)$, which  models the {\em long-term cost} of selecting node $i$ as the node that will execute the transaction as the sum of the frequency of accesses to the conflict classes in $S$ by every other node $j\neq i\in\Pi$:
$$LC(i,S)=\sum_{x\in S} \sum_{j\in\Pi\vee j\neq i}\mathcal{F}(j,x)$$
\noindent
where $\mathcal{F}(j,x)$ is defined as the per time-unit number of transactions
originated on node $j$ that have object $x$ in their dataset.
The first policy, which we term {\em long-term policy}, is obtained by setting the generic cost function $C(i,S)$ in (1) to $LC(i,S)$. This bases the decision on where to execute transactions on the statistics, collected over time, expressed by the frequencies $\mathcal{F}(j,x)$.

In order to derive the the \emph{short-term policy}, we first define the function $SC(i,S)$, which
expresses the immediate costs induced at the GCS level by different choices of where to execute a transaction:

\begin{equation*}
\label{eq:cost}
SC(i,S) = \left\{
\begin{array}{rl}
c_{URB} & \text{if } i=O \wedge \forall x\in S : \mathcal{L}(i,x)=1\\
c_{AB}+2c_{URB} & \text{if } i=O \wedge \exists x\in S : \mathcal{L}(i,x)=0\\
c_{p2p}+c_{AB}+2c_{URB} & \text{if } i\neq O \wedge \exists x\in S : \mathcal{L}(i,x)=0\\
c_{p2p}+c_{URB} & \text{if } i\neq O \wedge \forall x\in S : \mathcal{L}(i,x)=1
\end{array} \right.
\end{equation*}

\noindent
where the first case captures the fact that a single URB is required if the node that originated the transaction already owns all the leases required by it and the transaction is not forwarded;
the second case expresses the cost to request one or more leases on the node that originated the transaction, in case it does not own all required leases and the transaction is not forwarded;
the third case expresses the cost in a scenario in which a transaction is forwarded to a node that does not own all required conflict classes;
and the last case expresses the cost in a scenario in which the transaction is forwarded to a node that already owns the leases for all required conflict classes.

\begin{table}
\begin{center}
\begin{tabular}{|c|c|}
\hline
$\Pi$& the set of nodes in the system\\
\hline
$\mathcal{F}(i,x)$& access frequency from node $i$ to object $x$\\
\hline
$\mathcal{L}(i,x)$& equals 1 if node $i$ owns a lease on object $x$, 0 otherwise\\
\hline
$N_i$& $N_{i}=1$ if the transaction will be certified \\
& (after having acquired all leases) by node $i$, 0 otherwise\\
\hline
$C_{i,S}$ & the cost of selecting node $i$ to commit a tx.\\
&  that accesses the conflict classes in the set $leaselSet$\\
\hline
$c_{URB}$ & cost of performing URB\\
\hline
$c_{AB}$ & cost of performing AB\\
\hline
$c_{p2p}$ & cost of performing point to point communication\\
\hline
$O$ & the identity of the node that originated the transaction\\
\hline
$maxCPU$ & the maximum CPU utilization for a node\\
\hline
$CPU_{i}$ & the CPU utilization at node $i$\\
\hline
\end{tabular}
\end{center}
\caption{Parameters used in the ILP formulation.}
\label{table-LP}
\end{table}
The short-term policy is obtained by setting the generic cost function $C(i,S)$ in (1) to $SC(i,S)$.

It is easily seen that the ILP of Equation 1 can be solved in O($|\Pi|$) time regardless of whether the long-term or the short-term policy is used. The statistics required for the computation of the long-term policy are computed by gathering the access frequencies of nodes to conflict classes. This information is piggybacked on the messages exchanged to commit transactions/request leases. A similar mechanism is used for exchanging information on the CPU utilization of each node. For the short term policy, we quantify the cost of the P2P, URB and OAB protocols in terms of their communication-steps latencies (which equal 1, 2, and 3, respectively).


\section{Experimental Evaluation}
\label{sec:eval}

In this section, we compare the performance of {\thesystem}'s fine-grained leases and
transaction migration mechanisms with that of the baseline ALC protocol. Performance is evaluated using two benchmarks: a variant of the \emph{Bank} benchmark \cite{DBLP:conf/systor/CarvalhoRR11,HerlihyLM06} and the \emph{TPC-C} benchmark \cite{tpcc}.
We compare the following algorithms: ALC (using the implementation evaluated in
\cite{genrstm}), FGL (ALC using the fine-grained leases mechanism), MG-ALC (ALC extended with the transaction migration mechanism), and two variants of \thesystem (transaction migration on top of ALC using fine-grained leases), using the short-term (\thesystem-ST) and the long-term (\thesystem-LT) policies, respectively. Both ALC and \thesystem are implemented in Java and are publicly available~\cite{implementation}.

 All benchmarks were executed on a cluster of 4 replicas,
each comprising an Intel Xeon E5506 CPU at 2.13 GHz and 32 GB of RAM,
running Linux and interconnected via a private Gigabit Ethernet.

~\\
\noindent \textbf{Partitioned Bank Benchmark.} The \emph{Bank} benchmark \cite{DBLP:conf/systor/CarvalhoRR11,HerlihyLM06} is a well-known transactional benchmark that
emulates a bank system comprised of a large number of client accounts. We extended the Bank benchmark with support for various types of read-write and read-only transactions, for generating more realistic transactional workloads.
A \emph{read-write transaction} performs transfers between randomly selected pairs of accounts.
A \emph{read-only transaction} reads the balance of a set of randomly-selected client accounts.
Workloads consist of 50\% read-write transactions and 50\% read-only transactions of varying lengths.

We introduce data locality in the Bank benchmark as follows. Accounts are split into \emph{partitions} such that each partition is logically associated with a distinct replica and partitions are evenly distributed between replicas. A transaction accesses accounts of a single partition. A transaction originated on replica $r$ accesses accounts of a (randomly selected) partition associated with $r$ with probability $P$, and accounts from another (randomly selected) remote (associated with another replica) partition with probability $1-P$. Larger values of $P$ generate workloads that are characterized by higher data-locality and by smaller inter-replica contention. 

For the Bank application, the optimal migration policy is to forward each transaction $t$ to the replica with which the  partition accessed by $t$ is associated. We therefore implement and evaluate a third variant of \thesystem (called \thesystem-OPT) using this optimal policy.\footnote{Our MG-ALC implementation also uses this optimal migration policy.}

\begin{figure}[t!]
\begin{center}
        \includegraphics[scale=0.49]{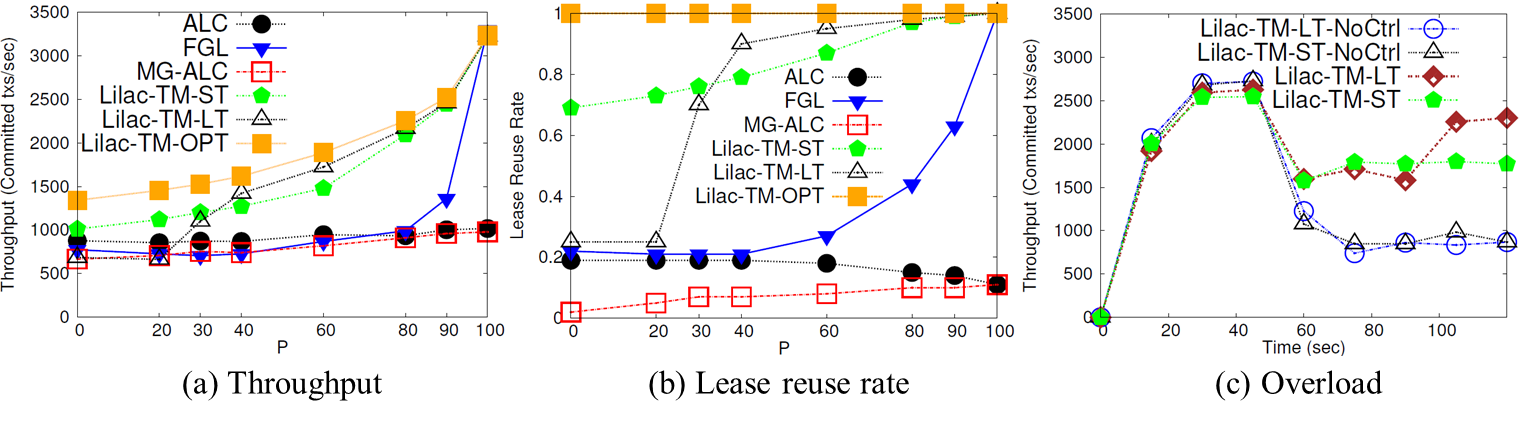}
\end{center}
\vspace{-1.5em}
\captionsetup{textfont=small}
\caption{Bank Benchmark}
\label{fig:bankBench}
\vspace{-1.25em}
\end{figure}

Figure \ref{fig:bankBench}(a) shows the throughput (committed transactions per second) of the algorithms we evaluate on workloads generated by the bank application with $P$ varying between 0\% to 100\%. 
We report in the following results obtained running 2 threads per node. Results using 4 threads per node are similar and are reported in the appendix.
 
Comparing ALC and FGL, Figure \ref{fig:bankBench}(a) shows that, while ALC's throughput remains almost constant for all locality levels, FGL's performance dramatically increases when locality rises above 80\%. This is explained by
Figure \ref{fig:bankBench}(b), that shows the \emph{Lease Reuse Rate}, defined as the ratio between
the number of read-write transactions which are piggy-backed on existing leases and the total number of read-write transactions.\footnote{Read-only transactions never request leases.} A higher lease reuse rate results in fewer lease requests, which reduces in turn the communication overhead and the latency caused by waiting for leases. FGL's lease reuse rate approaches one for high locality levels, which enables FGL and FGL-based migration policies to achieve up to 3.2 times higher throughput as compared with ALC and MG-ALC.

When locality is lower than 80\%, the FGL approach yields throughput that is comparable to ALC.
Under highly-contended low-locality workloads, FGL's throughput is even approximately 10\%-20\% lower than that of ALC.
This is because these workloads produce a growing demand for leases from all nodes.
FGL releases the leases in fine-grained chunks, which results in a higher load on $URB$-communication as compared with ALC.

The adverse impact of low-locality workloads on transaction migration policies, however, is much lower.
Migrating transactions to replicas where leases might already be present (or will benefit from acquiring it),
increases the lease reuse rate, which increases throughput in turn. Indeed, as shown by Figure \ref{fig:bankBench}(a), \thesystem achieves speed-up of between 40\%-100\% even for low-locality workloads (0\%-60\%) in comparison with ALC. For high-locality workloads, both FGL and \thesystem converge to similar performance, outperforming ALC by a factor of 3.2.

Comparing the performance of ALC and MG-ALC shows that using transaction migration on top of ALC does not improve the lease reuse rate as compared with the baseline ALC. This is because migration only helps if it is used on top of the fine-grained leases mechanism. The slightly lower throughput of MG-ALC in comparison to ALC is due to the overhead of the TF mechanism.

Next we evaluate the ability of \thesystem to cope with load imbalance. To this end, we set the benchmark to access with 20\% probability a single partition, $p$, from all the nodes, except for the single node, say $n$, associated with $p$, which accesses only $p$. In these settings, with all the considered policies, $n$ tends to attract all the transactions that access $p$. At second 40 of the test, we overload node $n$ by injecting external, CPU-intensive jobs. The plots in Fig.~\ref{fig:bankBench}(c) compare the throughput achieved by \thesystem with and without the mechanism for overload control (implementing Inequality (3)), and with both the long-term and the short-term policies. The data highlights the effectiveness of the proposed overload control mechanism, which significantly increases system throughput. In fact, the schemes that exploit statstics on CPU utilization (Lilac-TM-ST and Lilac-TM-LT) react in a timely manner to the overload of $n$ by avoiding further migrating their transactions towards it, and consequently achieve throughput that is about twice that of uninformed policies (Lilac-TM-ST-NoCtrl and Lilac-TM-LT-NoCtrl).


\vspace{-6pt}
~\\
\textbf{TPC-C.}
%
%
We also ported the TPC-C benchmark and evaluated {\thesystem} using it.
The TPC-C benchmark is representative of OLTP workloads and is useful to assess
the benefits of our proposal even in the context of complex workloads that
 simulate real world applications. It includes a wider variety of transactions
that simulate an application for a whole-sale supplier that supplies
\emph{items} (grouped in \emph{stocks}) from   a set of \emph{warehouses}
to \emph{customers} within sales \emph{districts}.
We ported two of the five transactional profiles offered by TPC-C,
namely the \emph{Payment} and the \emph{New Order} transactional profiles, that
exhibit high conflict rate scenarios and long running transactional workloads, respectively.
For this benchmark we inject transactions to the system by emulating a load balancer
operating according
to a geographically-based policy that forwards
requests on the basis of the requests' geographic origin: in particular requests sent
from a certain geographic region are dispatched to the node that is responsible for
the warehouses associated with the users of that region. To generate more realistic scenarios
we also assume that the load balancer can do mistakes by imposing that with probability
0.2 a request sent from a certain region is issued by users associated with warehouses that do not belong to that region.

\begin{wrapfigure}{RB}{0.4\textwidth}
\begin{center}
		\includegraphics[scale=0.35]{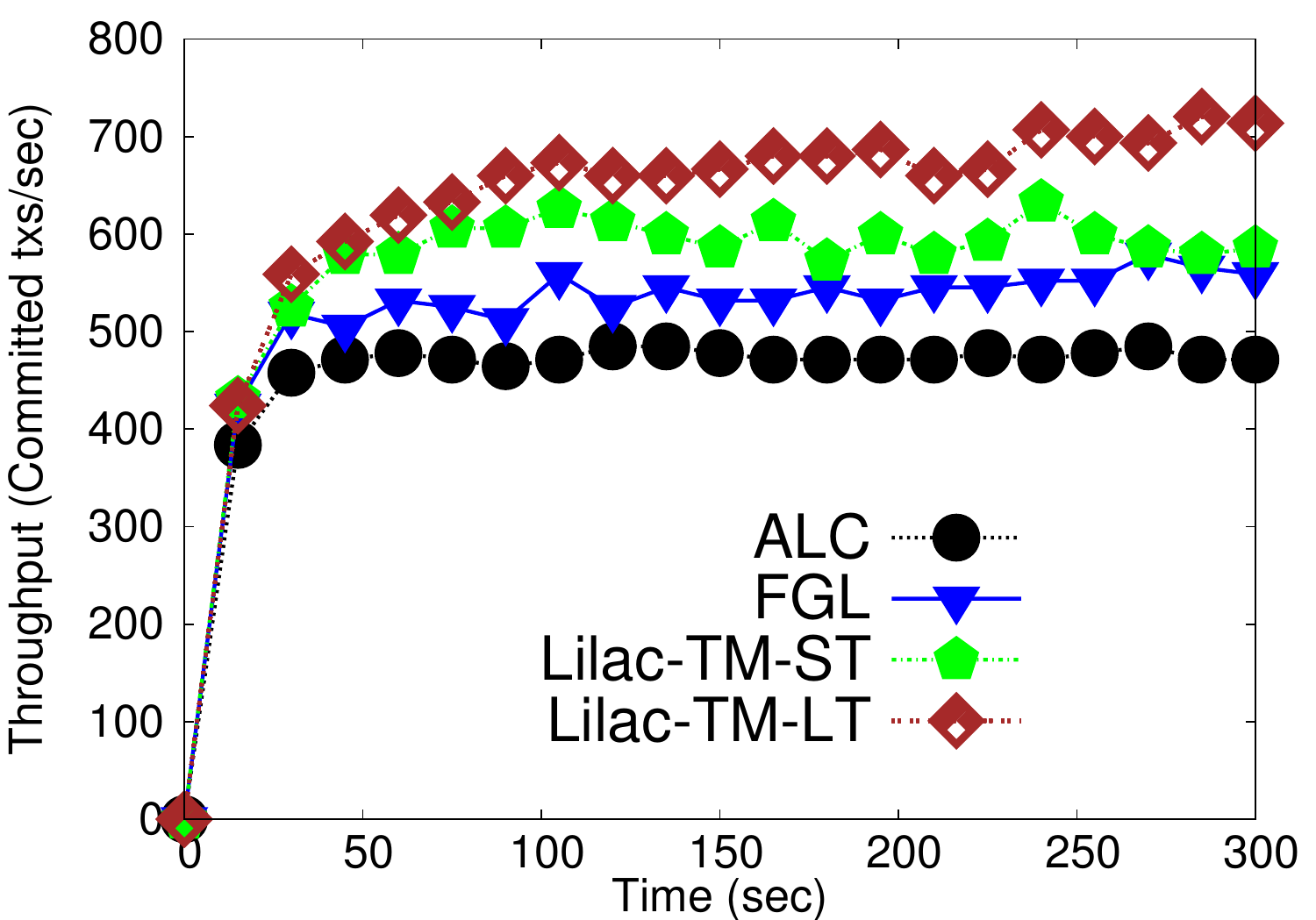}
\end{center}
\captionsetup{textfont=small}
\caption{TPCC}
\label{fig:tpccBench}
\end{wrapfigure}

In Figure~\ref{fig:tpccBench} we present the throughput
obtained by running a workload with 95\% Payment transactions and 5\% New Order
transactions; moreover in this case we show
the throughput varying over time in order to better assess the convergence of the reschedule policies.
We first notice that even in this complex scenario FGL
performs better than ALC due to better exploitation of the application,
and a higher leases reuse rate.
In addition, using the migration mechanism, driven by either the short term (ST) or the long term (LT) policy,
over FGL, achieves speedups of between 1.2 and 1.5
when compared to ALC. However, unlike the Bank Benchmark, in this case the ST
policy achieves only minor gains compared to the LT policy,
due to TPC-C's transactional profiles that generate more complex access patterns.
In fact, even when the data set is partitioned by identifying each partition as
a \emph{warehouse} and all the objects associated with that \emph{warehouse}, TPC-C's transactions
may access more than one partition. This reduces the probability that the ST policy can actually trigger a reschedule for
a transaction on a node that already owns all the leases necessary to validate/commit that transaction.
On the other hand the LT policy can exploit application locality thus
noticeably reducing lease requests circulation, i.e. the number of lease requests issued per second.

\section{Conclusions}
In this paper we introduced \thesystem, a fully decentralized, LocalIty-aware LeAse-based repliCated TM (\thesystem). \thesystem exploits a novel, self-optimizing lease circulation scheme that provides two key benefits: (1) limiting the frequency of lease circulation, 
and (2) enhancing the contention management efficiency, by increasing the probability that conflicting  transactions are executed on the same node.

By means of an experimental evaluation based on both synthetic and realistic, complex benchmarks, we have shown that \thesystem can yield significant speed-ups, reaching peak gains of up to 3.2 times with respect to state of the art lease-based replication protocols.

\bibliographystyle{abbrv}
\bibliography{stm,cross,macros,transactional-replication}

\newpage
\appendix

\section{Additional experimental results}

In this section we present evaluation results obtained using 4 threads per node with both the Bank benchmark and the TPC-C benchmark, omitted from the main body of the paper.

We start by analyzing the results of the Bank benchmark.
Although the results are fairly similar with those obtained using two threads per node, there are nevertheless two noticeable differences. First, it can see that when locality exceeds 80\%, both \thesystem and FGL enjoy a higher speed-up (in comparison with the baseline ALC) for 4 threads as compared with 2 threads.

As described in the main body of the paper, unlike \thesystem (or FGL), in ALC threads spend much of their time obtaining leases even in high locality scenarios. In fact, even though in these scenarios nodes globally own leases associated with most of the data set accessed by their transactions, they cannot successfully reuse the leases across transactions, due to the coarse granularity of the lease management scheme employed by ALC.  Indeed, in high locality scenarios the achieved speed-up is even higher than for two threads, obtaining up to 4.8 times higher throughput with respect to ALC and MG-ALC.

The second difference is that when locality is under 80\%, 4 threads are better able to compensate for the $URB$-communication overhead caused by releasing the leases in fine-grained chunks.
While only some of the threads pay the price of requesting the leases, the others are able to simply piggy-back the acquired leases and commit, thus eliminating the gap between FGL and ALC and increasing the speed-up for \thesystem.

For the TPC-C benchmark we show the throughput (committed transactions per second) obtained by running the same workload adopted for the configuration of 2 threads per node (i.e. 95\% of Payment transactions and 5\% of New Order transactions). Also in this case, it is noticeable that the combination of the migration mechanism and the fine-grain lease management approach outperforms the baseline ALC, achieving a maximum speed-up of 1.5 when adopting the long-term policy. However, this is not the case for the short-term policy, which performs even worse than the baseline. This stems from the fact that the probability to actually find all the necessary leases to validate/commit a transaction during a migration is very low due to the higher logical contention. This means that even if the policy considers a node N as a good candidate for the migration of a transaction T (because it forecasts that N will have all the leases necessary to validate/commit T), the execution of T on N could be followed by an additional lease request since N may not have the required leases anymore because these were already granted to other nodes.

\begin{figure}
\begin{center}
	\subfloat[Bank]{
		\includegraphics[scale=0.37]{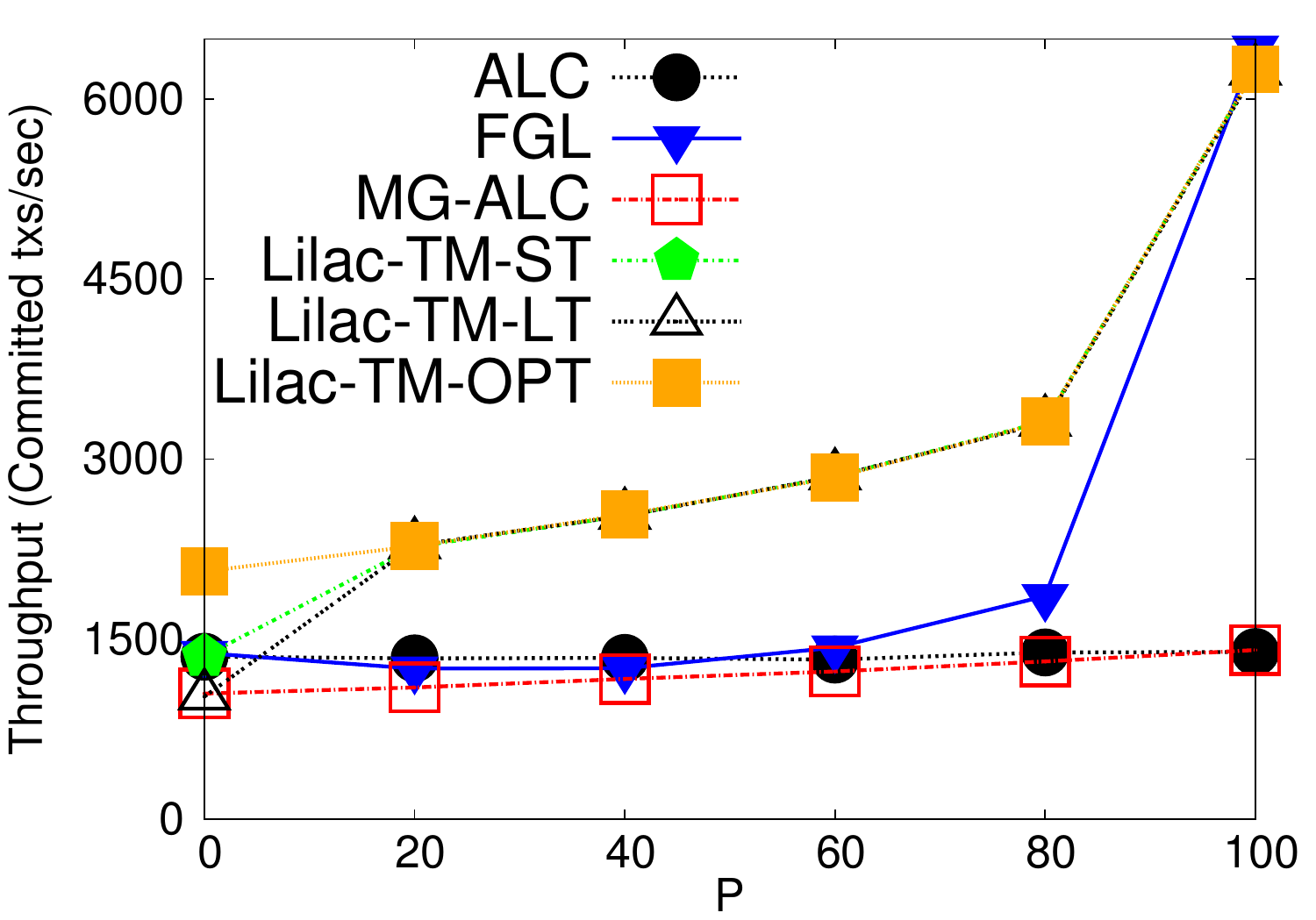}
	}\hspace{0.48em}
	\subfloat[TPCC]{
		\includegraphics[scale=0.37]{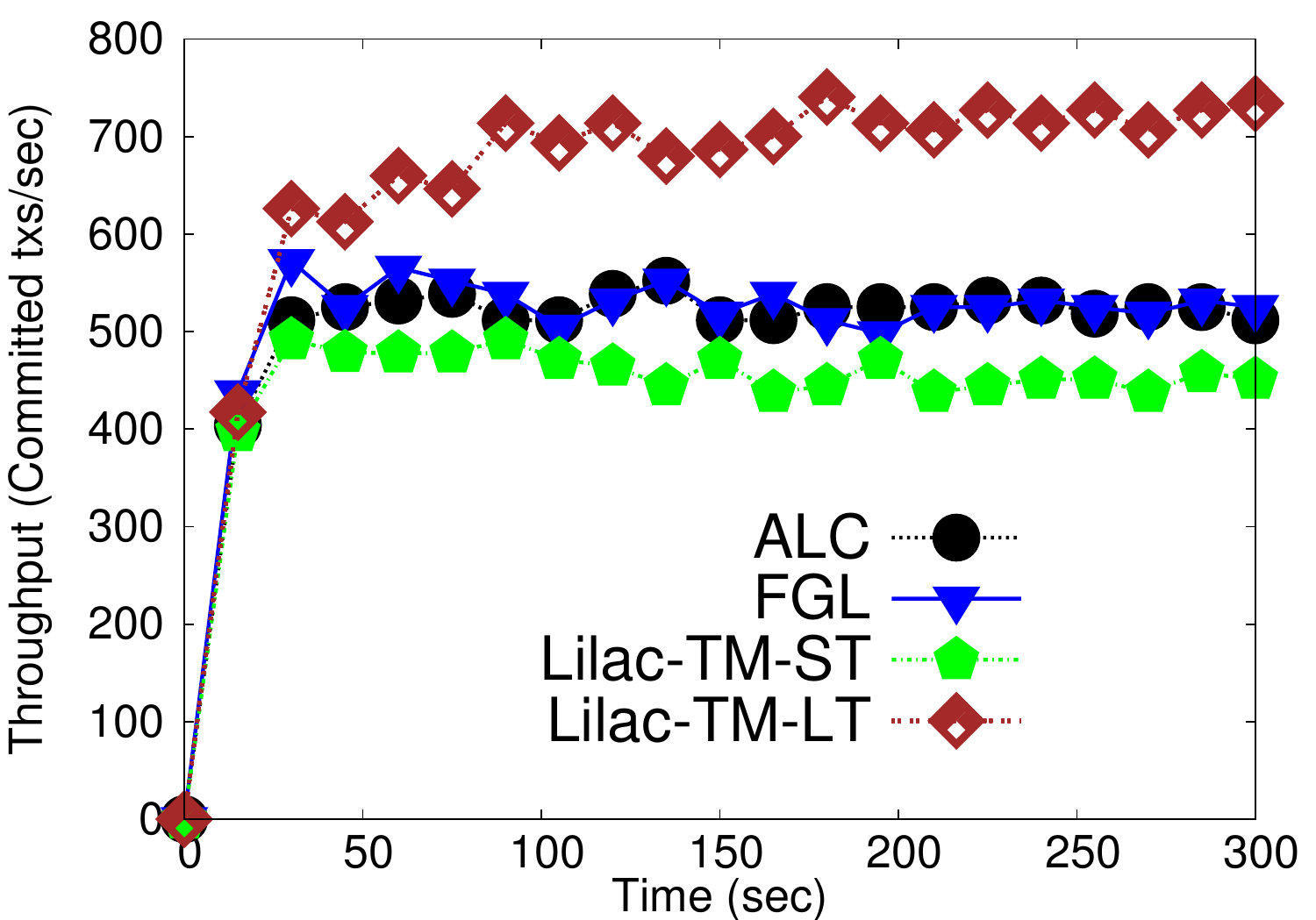}
	}
\end{center}
\vspace{-1.5em}
\captionsetup{textfont=small}
\caption{4 threads per replica}
\label{fig:4threads}
\vspace{-1em}
\end{figure}

\end{document}